# TUNABLE INTEGRATED-OPTICS NANOSCALED DEVICES BASED ON MAGNETIC PHOTONIC CRYSTALS

*M. Vasiliev[1], V.I. Belotelov[2], K.E. Alameh[1], R. Jeffery[1], V.A. Kotov[3], A.K. Zvezdin[3]*

1 - Centre for MicroPhotonic Systems, Electron Science Research Institute
Edith Cowan University, Joondalup, WA, 6027, Australia.
2 - M.V. Lomonosov Moscow State University, 119992, Leninskie gori, Moscow, Russia
3 - A.M. Prokhorov General Physics Institute, RAS, 119991, Vavilov St., 38, Moscow, Russia

## ABSTRACT

Magnetooptical properties of magnetic photonic crystals have been investigated in the view of their possible applications for the modern integrated-optics devices. A "transfer matrices" formalism was expanded for the case of oblique light incidence on the periodic nanoscaled magnetic multilayered systems. Several new effects such as the Faraday effect dependence on the incidence angle and the tunability of the bandgap defect modes spectral location by external magnetic fields were found. Several possibilities of one-dimensional magnetic photonic crystals applications for the optical devices are discussed. Initial steps towards the practical implementation of the proposed devices are reported.

## 1. INTRODUCTION

For the last decade, photonic crystal (PC) materials have been the subject of intense theoretical and experimental studies, which is largely due to their outstanding optical properties that are very promising for the numerous integrated-optics applications [1,2]. Photonic crystals are nanostructured materials with one-, two-, or three-dimensional periodicity in their dielectric constant. One of the most prominent features of PCs is the presence of a photonic band-gap – the spectral range for which light propagation through the photonic material is completely prohibited.

Investigation of PCs with paramagnetic or ferromagnetic constituents - magnetic photonic crystals (MPC) is of prime importance nowadays. Magnetic materials placed periodically in the PC system lead to several new optical phenomena and give rise to the substantial enhancement of conventional magnetooptical (Faraday, Kerr, Voight) effects [3-5]. This makes MPC even more promising materials for the development of modern nanoscaled optical devices.

One dimensional (1D) MPC usually represents alternating magnetic and nonmagnetic layers of a quarter-wavelength thickness [4,5]. However, some structural defects in that periodicity can produce additional transmission and Faraday rotation resonances inside the photonic band-gap. The examples of two-dimensional MPCs are periodical hole arrays in the transparent dielectric matrix filled with a magnetic medium. Systems of periodically located spheres with voids occupied by some magnetic fluids are common examples of three-dimensional MPCs [6].

This work is devoted to the investigation of 1D MPCs in the near-infrared wavelength range. We describe a theoretical approach for the calculation of their optical properties and consider several possible applications for the modern nanoscaled optical devices. First steps of their experimental implementation are also presented.

## 2. THEORY

There are several different approaches applicable for the calculation of the multilayered systems optical properties [5,7,8]. Among them, the 4×4 transfer matrices method was proven to be an efficient and intuitively clear technique [5,9]. That is why in this work we adopt this approach for the modeling of MPCs.

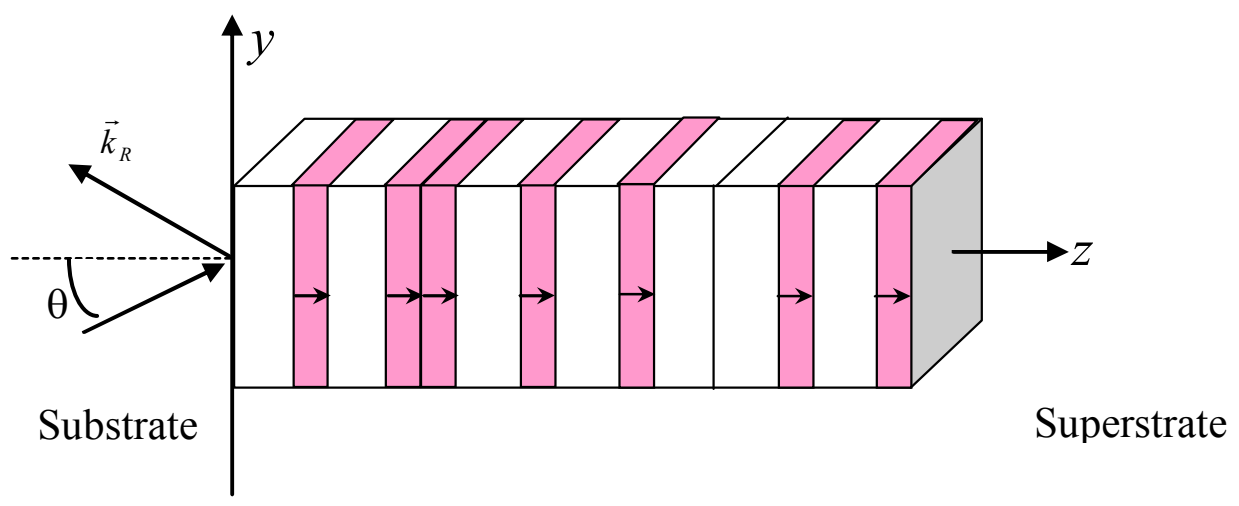

Fig.1. The geometry of the problem. The MPC structures of $(NM)^2(MN)^3(NM)^2$ with two defects is shown.

The essence of the transfer matrices method applied for the simulation of 1D MPCs is as follows: Let us consider a 1D MPC placed between substrate of dielectric constants $\varepsilon_a$ and superstrate of dielectric constant $\varepsilon_b$ (Fig.1). The coordinate system is chosen in such a way that *z*-axis is perpendicular to the interfaces of MPC layers. Visible or near-infrared electromagnetic radiation is incident at the angle θ. During its propagation through the structure, multiple reflections and refractions take place producing a relatively complex distribution of the electromagnetic field amplitudes inside MPC. Due to the periodicity of the MPC system, the propagating field is represented by Bloch waves. Nevertheless, for further analysis in the framework of transfer matrices, it is much more convenient to consider the electromagnetic field in each MPC layer as a superposition of the so-called "proper modes", which in the 1D case are four waves propagating independently of each other in the two opposite directions and having the states of polarization which are preserved during propagation.

The main purpose of the transfer matrices method is to relate the electromagnetic fields within the substrate and the superstrate. In order to accomplish this, one needs to describe fully the proper modes propagation throughout the entire system. That is why three basic matrices are introduced for each layer. These are proper modes column vector $A_n$, propagation matrix $P_n$, and dynamic matrix $D_n$. The column vector $A_n$ gives the electric field amplitudes of proper modes at the exit boundary of the *n*-th layer, the propagation matrix $P_n$ represents the phase shifts of the proper modes throughout the *n*-th layer, and the dynamic matrix $D_n$ relates the amplitudes of the proper modes in adjacent layers.

The boundary conditions for the tangential components of electric and magnetic fields at the boundary between $n$-th and $(n+1)$-th layers lead to the following equation for the amplitudes of proper modes in the superstrate and substrate media:

$$A_{N+1} = D_{N+1}^{-1} D_N P_N D_N^{-1} \ldots D_1 P_1 D_1^{-1} D_0 A_0 \quad (1)$$

Once proper modes amplitudes before and after MPC are related, one can find the electric field amplitudes for the reflected and transmitted radiation, which is necessary for the optical and magnetooptical properties calculation. For example, provided that p-polarization is incident, the Faraday rotation angle is determined by

$$\Phi = \frac{1}{2}\operatorname{atan}\left(\frac{2\operatorname{Re}\chi}{1-|\chi|^2}\right), \quad (2)$$

where $\chi = E_s/E_p$, $E_s$, $E_p$ are proper modes amplitudes of the transmitted light.

One of the advantages of the transfer matrices technique is its universality, which enables the optical properties of both periodic and non-periodic multilayered structures to be accurately modeled.

Thus, the most important point in transfer matrices approach is to find proper modes amplitudes in each layer. They can be determined from the Fresnel equation:

$$n^2\vec{E} - \vec{n}(\vec{n}\vec{E}) = \hat{\varepsilon}\vec{E} \quad (3)$$

where $\vec{n} = c\vec{k}/\omega$, $\vec{k}$ is the wave vector, and $\hat{\varepsilon}$ is the dielectric tensor of the medium. Since the tangential component of wave vector is continuous at all interfaces, then $n_y = \sqrt{\varepsilon_a}\sin\theta$. The component $n_z$ is determined from the condition of solvability of the equation (3) with respect to the electric field components.

There are infinite number of the proper modes bases for the electromagnetic waves in non-magnetic layers, but usually *s*- and *p*- plane polarizations are chosen. The unit vectors of electric and magnetic fields for s- and p-polarizations of light with $k_z > 0$ are given by

$$\vec{e}_s = (1,0,0), \quad \vec{e}_p = \left(0, \frac{n_z}{\sqrt{\varepsilon}}, -\frac{n_y}{\sqrt{\varepsilon}}\right) \quad (4)$$

For magnetic layers the situation is different due to the presence of the gyration vector $g$. In general, they are to be characterized in terms of the permittivity and permeability tensors $\hat{\varepsilon}$ and $\hat{\mu}$, but at the visible and near-infrared wavelength range the latter can be considered as the unit tensor.

The permittivity tensor of an optically isotropic magnetic medium magnetized along the *Z*-axis has the following non-zero components [10]:

$$\varepsilon_{11}^{(m)} = \varepsilon_{22}^{(m)} = \varepsilon_{33}^{(m)} = \varepsilon_m, \; \varepsilon_{12}^{(m)} = -ig, \; \varepsilon_{21}^{(m)} = ig \quad (5)$$

It is assumed here that the second-order magnetooptical effects are negligibly small for the considered materials.

Using (4) and (5), one can get the following expression for the longitudinal wavenumber $n_z$ in a magnetic layer magnetized along *Z*-axis

$$\left(n_{z\pm}^{(m)}\right)^2 = \varepsilon_m - n_y^2 \pm g\sqrt{\left(1 - \frac{n_y^2}{\varepsilon_m}\right)}. \quad (6)$$

The proper modes electric field components are determined from the Fresnel equation (3):

$$E_{x,\pm} = \frac{-ig}{\left(n_\pm^{(m)}\right)^2 - \varepsilon_m} E_{y,\pm}, \quad E_{z,\pm} = \frac{n_y n_{z\pm}^{(m)}}{n_y^2 - \varepsilon_m} E_{y,\pm}, \quad (7)$$

where $n_\pm^{(m)} = \sqrt{n_y^2 + \left(n_{z\pm}^{(m)}\right)^2}$.

It is important to note here that because of the relatively small values of the gyration $g$ ($g \sim 10^{-4} \div 10^{-2}$), the condition $\varepsilon_m \cos\theta_m >> g$ is satisfied for almost all angles $\theta_m$, where $\theta_m$ is the average angle between the *Z*-axis and proper mode wave vectors inside the magnetic medium. Taking these conditions into account, expressions (6) and can be simplified substantially, giving

$$n_{z\pm}^{(m)} = \sqrt{\varepsilon_m}\cos\theta_m \pm \frac{g}{2\sqrt{\varepsilon_m}}. \quad (8)$$

It can be shown that for the case of oblique incidence, the proper modes inside the magnetic medium are two elliptically polarized waves with the electric field vector circumscribing cone surfaces with the elliptical basis and opening angle very close to 180º, so the cone surfaces are very close to planes. Consequently, to describe the Faraday effect for the oblique incidence one can ignore slight non-orthogonality of the electric field and the wave vector and consider two mutually perpendicular p- and s-directions, in analogy with the conventional terminology.

The Faraday effect is determined by the value of the phase shift between two orthogonal elliptical polarizations, which is proportional to $\Delta k$, which, taking into account (8), is given by

$$\Delta k = \pi/\lambda_0\left(n_{z+} - n_{z-}\right) = \frac{\pi g}{\lambda_0\sqrt{\varepsilon_m}}. \quad (9)$$

where $\lambda_0$ is the incident light wavelength in the vacuum. That is why for relatively small medium magnetization the Faraday rotation can be assumed independent of the incidence angle and, hence, light ellipticity can be neglected. Consequently, the change in Faraday rotation with respect to the incidence angle is mainly due to the redistribution of electromagnetic wave energy among magnetic and nonmagnetic layers of MPC.

Once the proper modes for magnetic and nonmagnetic constituents are found, it is possible to calculate transfer matrices $A_n$, $P_n$, $D_n$ and obtain transmittance, reflectance and Faraday rotation spectra using (1) and (2).

## 3. RESULTS AND DISCUSSION

Due to the magnetooptical effects, MPC structures can be used as tunable optical nano-devices. This can be accomplished by two approaches, namely, (i) the presence of the Faraday effect inside MPCs allows for the substantial polarization rotation, which depends on whether s- or p-polarised wave is incident; and (ii) magnetooptical light-medium interaction leads to the changes in s- and p-waves transmittance spectra.

During the modelling of the MPCs optical properties, we have found that they are considerably more influenced by the magnetic field at the wavelengths of bandgap defect modes, rather than near the band-gap edges. For this reason, the process of engineering nano-devices based on MPC involves the optimisation of quasi-periodic multilayer structures and careful placement of structural defects, which can generate bandgap modes with desired properties. We have developed efficient C++/Windows-based algorithms capable of assisting in the MPC design process [11]. In the remaining parts of this paper, we present our on-going work on the development of several novel nano-level optical devices for applications in the telecommunications industry, which are based on the unique properties of MPC structures optimised for operation in the oblique incidence geometry. The features of the oblique-incidence MPC operation we found most remarkable are the tunability of the overall spectral region of operation achieved by controlling the angle of incidence, and the tunability of the spectral locations of bandgap defect modes resonances achieved by changing the magnitude of the applied external magnetic field.

### 3.1. WDM demultiplexers with transmission loss equalisation

The ability to tune the operational wavelength of the MPC by varying the angle of incidence opens up a possibility of designing demultiplexers with magneto-optic equalization of transmission loss. A passive or reconfigurable (based on a spatial light modulator utilizing liquid crystal on silicon (LCOS) technology – an Opto-VLSI processor) diffraction grating can be used to generate an appropriate (peak transmission-tuned) array of incidence angles for the light carrying several WDM signal channels [12]. An Opto-VLSI processor is an array of liquid crystal (LC) cells whose crystallographic orientations are independently addressed by a Very-Large-Scale-Integrated (VLSI) circuit to create a reconfigurable, reflective, holographic diffraction grating plate. Application of voltage between the electrodes of the VLSI circuit induces a phase hologram in the LC layer, resulting in optical beam steering and/or beam shaping. Fabricated Opto-VLSI devices are electronically controlled, software-configured, polarisation independent, cost effective because of the high-volume manufacturing capability of VLSI and very reliable since beam steering is achieved with no mechanically moving parts. Fig. 2 shows a typical layout of an Opto-VLSI processor. Also shown is typical LC cell design. Usually Indium-Tin Oxide (ITO) is used as the transparent electrode, and evaporated aluminium is used as reflective electrode. Opto-VLSI processors can generate stepped blazed grating for optical beam steering, as well as multicasting grating for arbitrary beam splitting, where the diffraction orders are deliberately enhanced to generate an arbitrary beam splitting profile [13].

The design of a WDM equaliser employing an Opto-VLSI processor as an angular demultiplexor and a MPC is shown schematically in Fig. 3(a). The light incident on the MPC has to be polarized (in an arbitrary plane for the case of near-normal incidence, since the s- and p-polarization responses in both transmission and Faraday rotation are then nearly identical). Fig. 3(b) shows typical input and equalized WDM spectra.

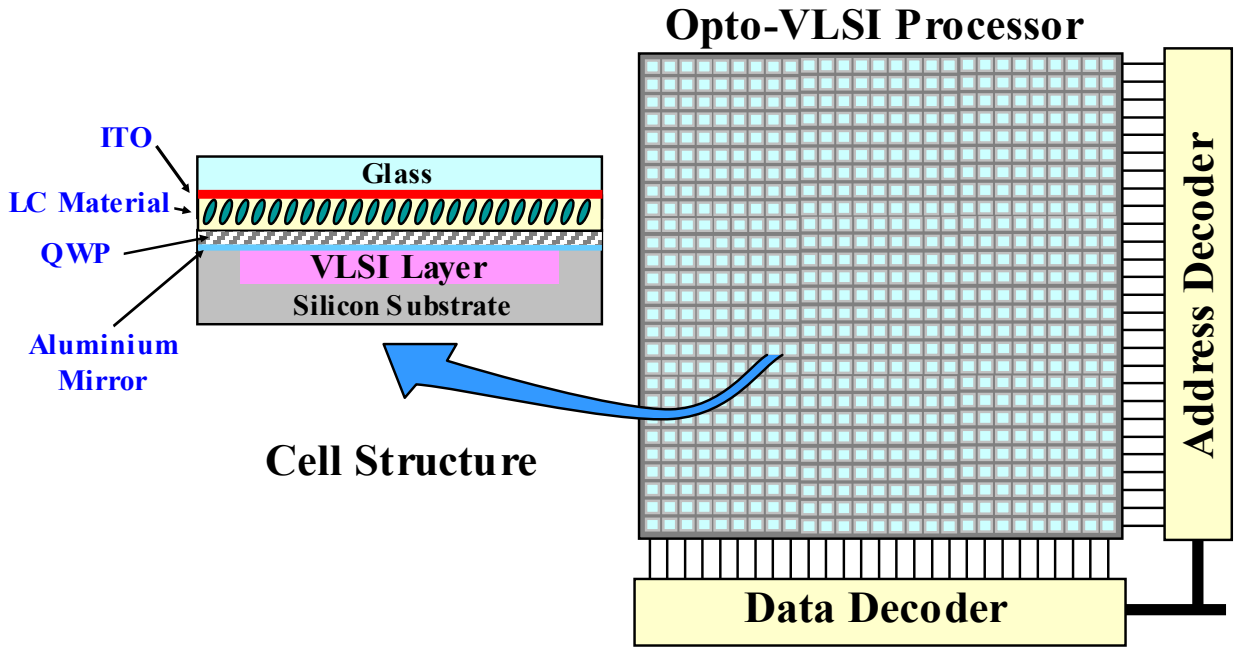

Fig. 2. Typical Opto-VLSI processor and an LC cell structure design.

Another polarizer (analyzer) can be used in the optical path after the MPC to achieve extinction when the plane of polarization is rotated magnetooptically by 45 degrees in a chosen direction. The transmitted intensity in each channel can then be controlled independently by rotating the required planes of polarization by ±45 degrees with magnetic fields applied locally using an array of integrated current-carrying coils (deposited onto MPC exit surface) that encircle the propagation paths of light. The crosstalk in the output will be very low due to the spatial separation of the channel propagation paths and the spectral sharpness of the MPC transmission resonances which are tuned to their corresponding angles of incidence. Due to the very low absorption of commonly used magnetic layer materials (bismuth or cerium-substituted YIG) in the 1.55 μm optical telecommunications window, it is possible to design MPC structures possessing very sharp spectral resonances having high transmission coupled with significantly enhanced Faraday rotation. In this case, relatively small magnetic fields can achieve large Faraday rotations within narrow spectral windows, which can be tuned by changing the incidence angle. For example, a MPC structure with a design formula $(NM)^{12}(MN)^{1}(NM)^{1}(MN)^{12}$, in which $SiO_2$-layers are used for non-magnetic constituent ( $\varepsilon = 2.24$ ), has a very sharp resonant behaviour in both transmission and Faraday rotation near the wavelength of 1.55 μm. This quarter-wave stack structure is composed of 52 layers and has a thickness of 11.29 μm. The spectral response of this structure, its variation with the angle of incidence and its optical response to the gyration are shown in Fig. 4. Here, we assumed two types of the surrounding media pairs, namely, (i) air and glass as before ( $\varepsilon_a = 1, \varepsilon_b = 2.310$ ), and (ii) the substrate and the exit medium which are index-matched to the mean refractive index of the structure ( $\varepsilon_a = \varepsilon_b = 3.168$ ).

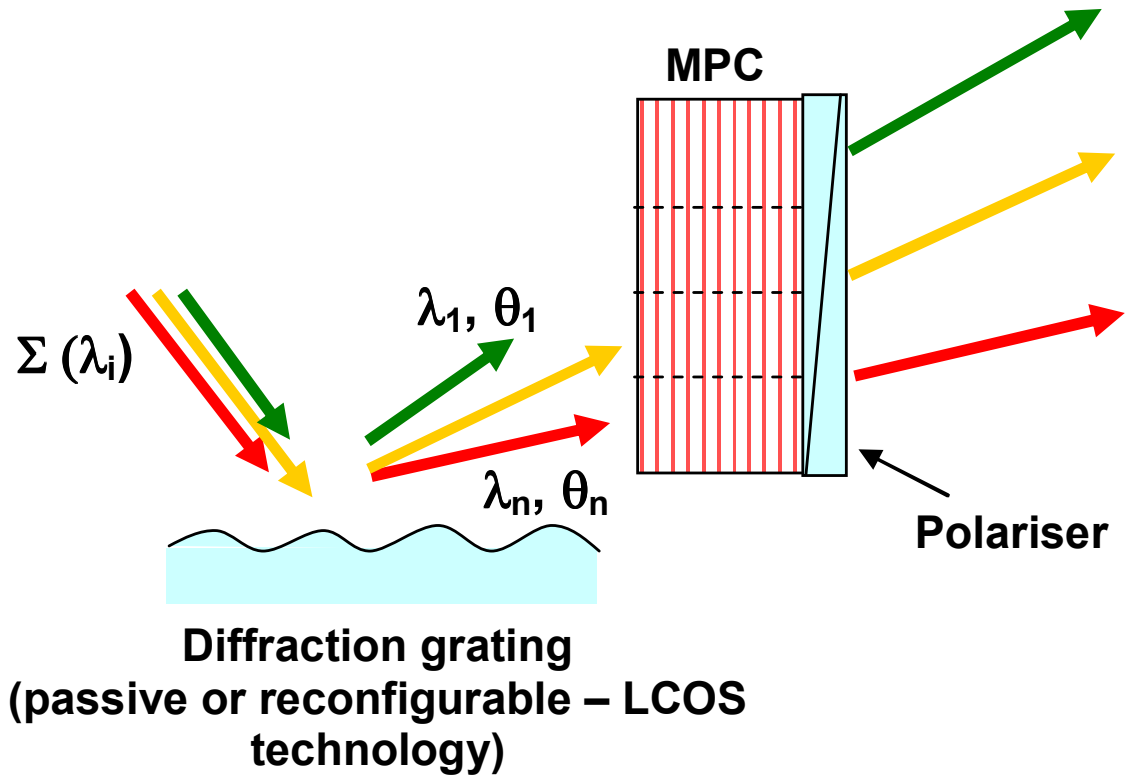

(a)

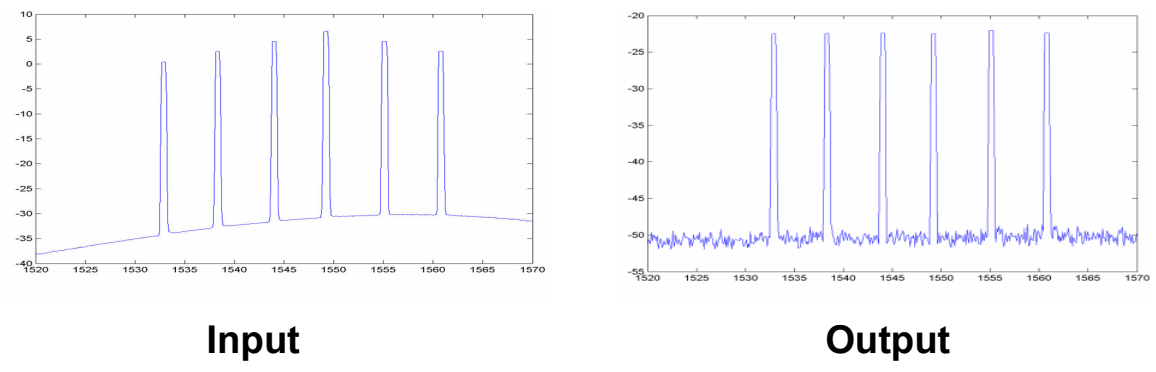

(b)

Fig. 3. (a) Principal diagram of MPC-based WDM equalizer utilising the oblique incidence geometry for localised control of channel transmittance, (b) input and equalized WDM spectra.

The gyration required for achieving the 45 degrees of rotation is $g$ = 0.00035, which can be achieved at about 3.8 % of the saturation magnetisation for Ce:YIG.

The splitting of two opposite elliptical polarizations resonances and its associated reduction in peak transmission observed at high levels of induced gyration limit the dynamic range of the device and introduce some extra loss, however, in some equalizer applications these drawbacks can be outweighed by the advantages of the high-speed operation characteristic of the magneto-optic devices. The response time required for switching of each channel from a minimum to a maximum of transmitted intensity can be as short as tens of nanoseconds.

The range of incidence angles (and, consequently the number of channels that can be processed by a single device of this type) can in principle be extended to large angles if control of the polarization of incident light is implemented. The wavelength-tuning curve (the dependency of the MPC operational wavelength on the angle of incidence) of the structure being considered is shown in Fig.5. This dependency is highly nonlinear for very small angles of incidence, but for moderate incidence angles, it can be approximated by a linear function with a slope of about 1 nm/deg for index-unmatched air and glass surrounding media and 6.7 nm/deg for index-matched surrounding media. The structure of the dispersion grating to be used for demultiplexing the multi-channel optical input must be optimized to closely match its angular dispersion function with a selected section of the wavelength-tuning curve of the MPC.

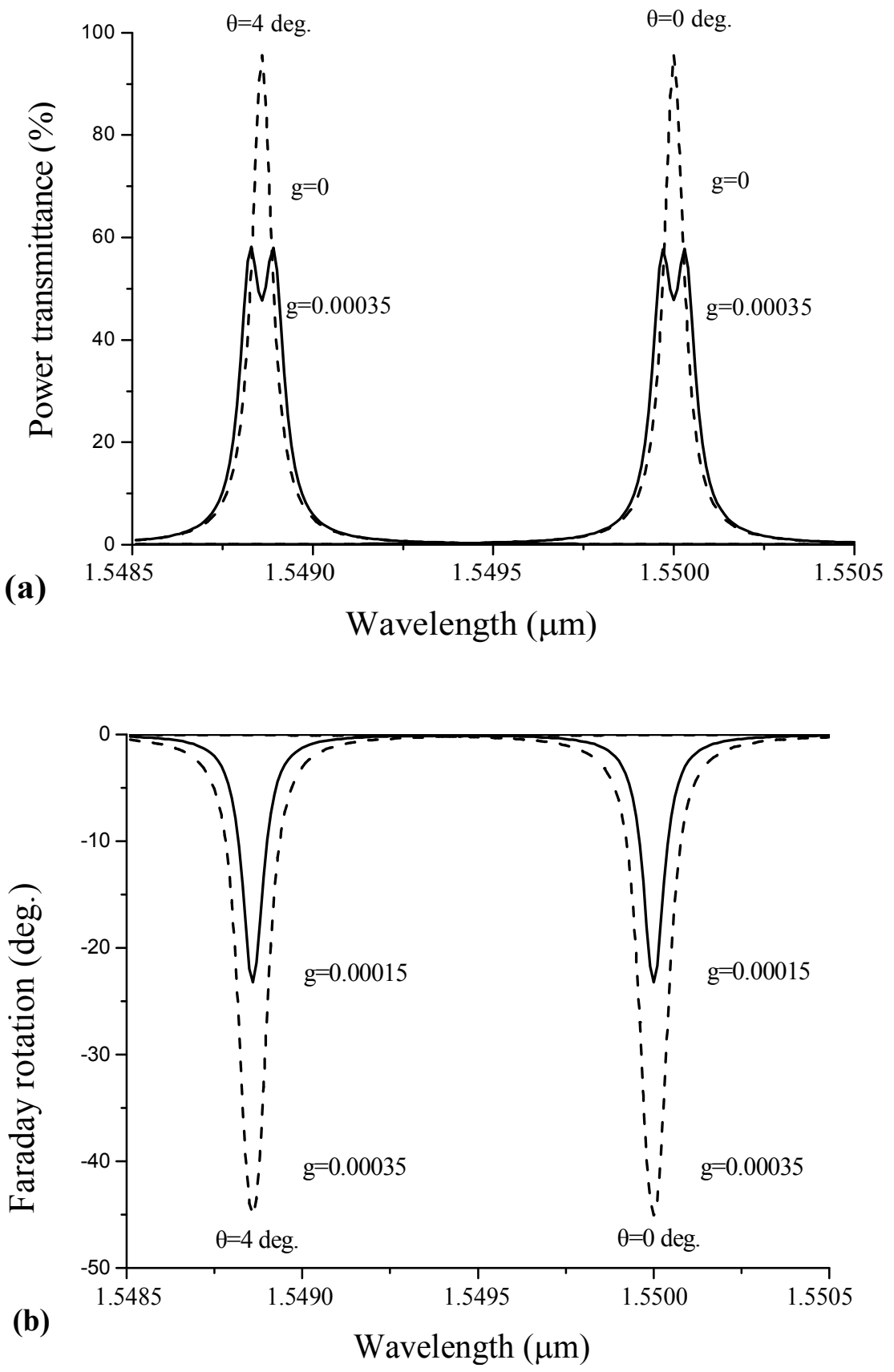

Fig. 4. Transmittance and Faraday rotation spectra of the structure $(NM)^{12}(MN)^{1}(NM)^{1}(MN)^{12}$ at normal and 4 degrees incidence with various strengths of the applied magnetic field. (a) Averaged (unpolarised) transmittance spectra at normal and 4º incidence and their variation with gyration; (b) Faraday rotation spectra for p-component at normal and 4º incidence at various gyrations. The spectral responses of s- and p-components of polarization are almost identical at near normal incidence. $\varepsilon_a = 1, \varepsilon_b = 2.310$.

An array of surface-deposited microcoils of 100 μm internal loop diameter and having either one or three turns was designed, fabricated and tested for its suitability for the proposed equalizer application. During our experiments, a single-layer, 20-μm thick film of Bi-substituted YIG was used to confirm the viability of achieving sufficient levels of magnetisation within the film for the operation of proposed device using a single-turn microcoil. The geometry of microcoils and a schematic of our experimental set-up are shown in Fig. 6. A current pulser circuit was designed to output pulses of 2 μs duration at a repetition rate of 1 kHz which enabled currents of up to 6 A to be used during testing. The magneto-optic response was also studied using DC currents of up to 0.3 A for field generation.

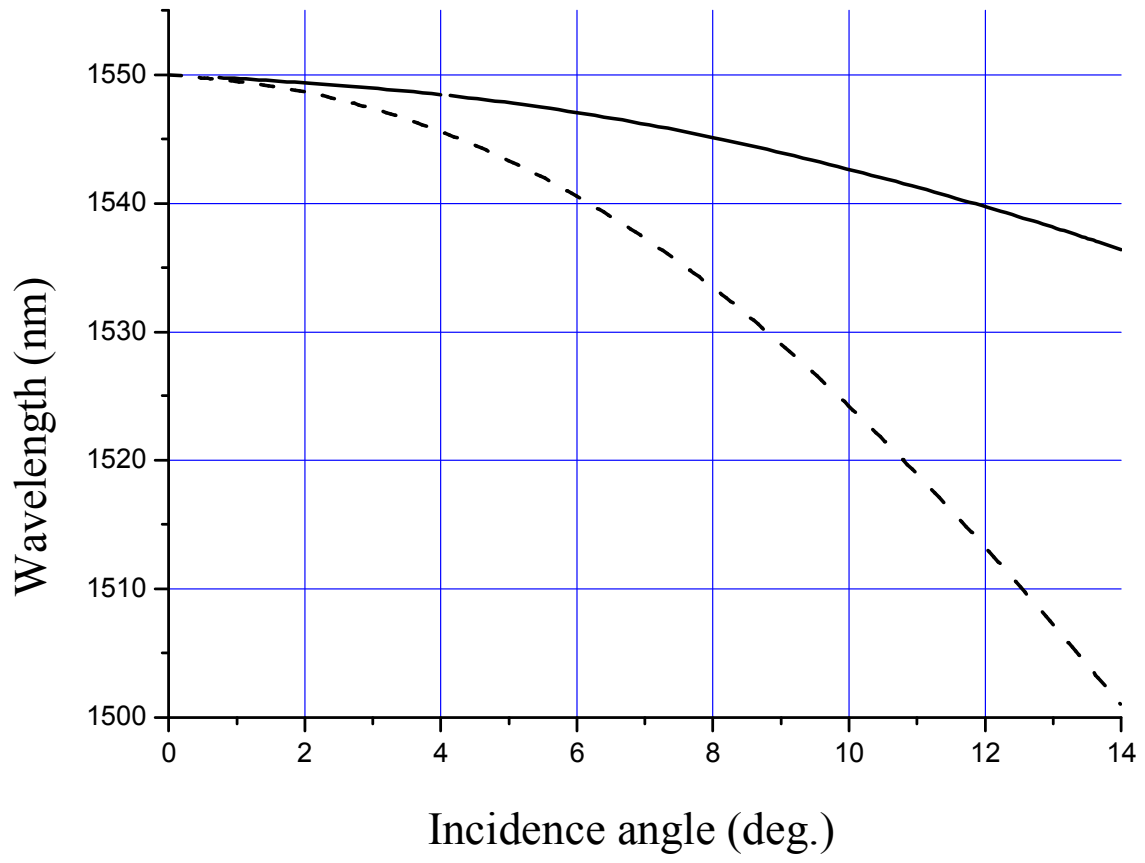

Fig. 5. MPC operational wavelength as a function of incidence angle for the structure $(NM)^{12}(MN)^{1}(NM)^{1}(MN)^{12}$ sandwiched between (i) air and glass ($\varepsilon_a = 1, \varepsilon_b = 2.310$) (solid line), or (ii) two identical media index-matched to the mean refractive index of the structure ($\varepsilon_a = \varepsilon_b = 3.168$) (dashed line).

The investigations revealed that the saturation of magnetisation in the film was reached at a coil current of about 2.8 A, confirming the suitability of our system for the proposed application with both DC and pulsed currents, provided that the optimised MPC structure requires less than about 10% of the saturation magnetisation to achieve Faraday rotations approaching 45º.

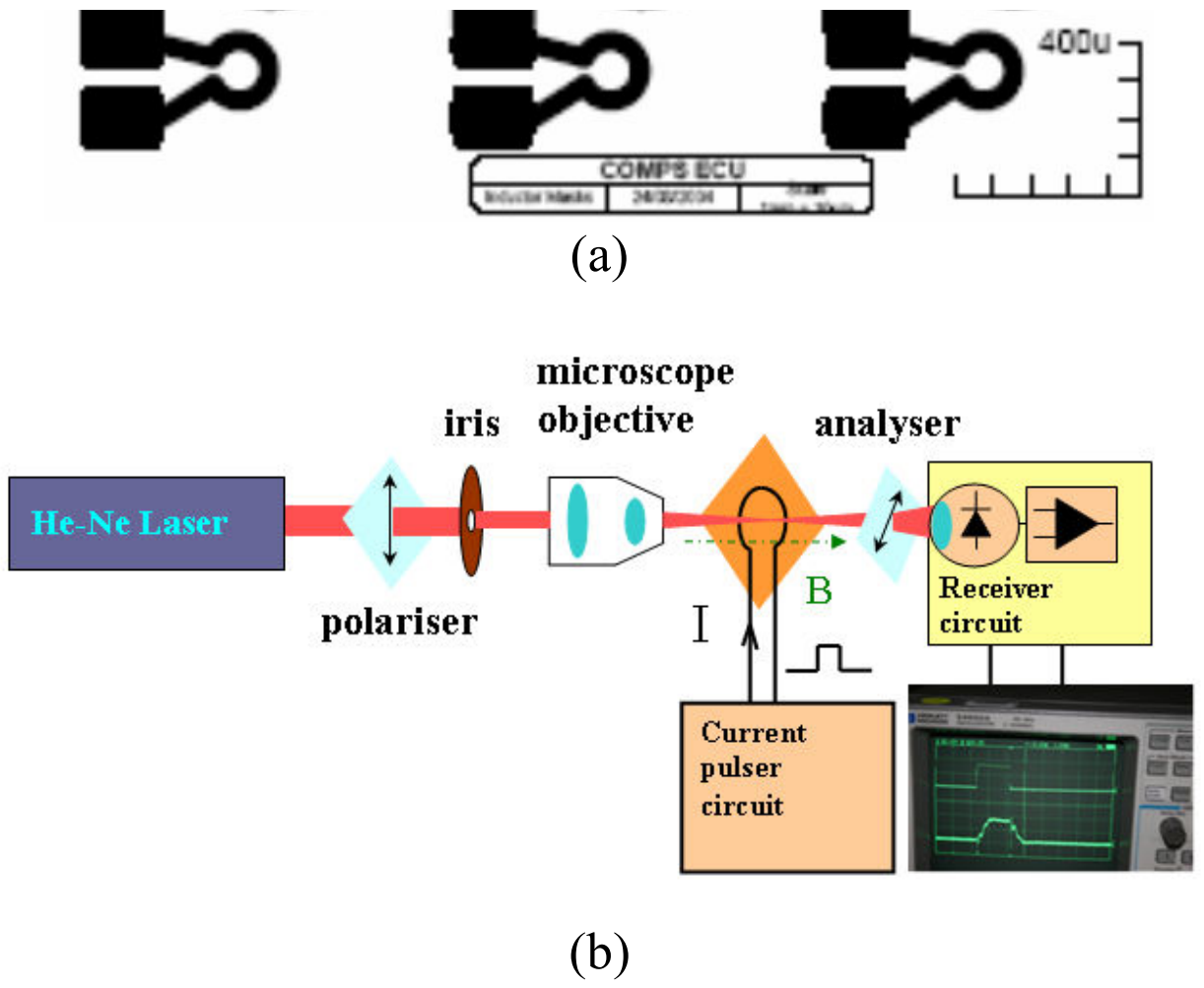

Fig. 6. (a) Geometry of microcoils used for magnetic field generation; (b) Schematic of the experimental set-up used for characterisation of magneto-optic response and studies of magnetisation dynamics.

### 3.2. Tunable optical switches employing magnetic control of bandgap defect modes.

Medium magnetization also affects the wavelength of the transmittance peak, as shown in Fig. 7 where transmittance spectra of the incident p-wave incident onto the MPC of formula $(NM)^{31}(MN)^{18}(NM)^{31}$ for $g = 0$, $g = 0.01$, $g = 0.03$, and $g = 0.05$ are plotted. The wavelength shift transmittance peak is proportional to the medium gyration, and, consequently, to the medium magnetization.

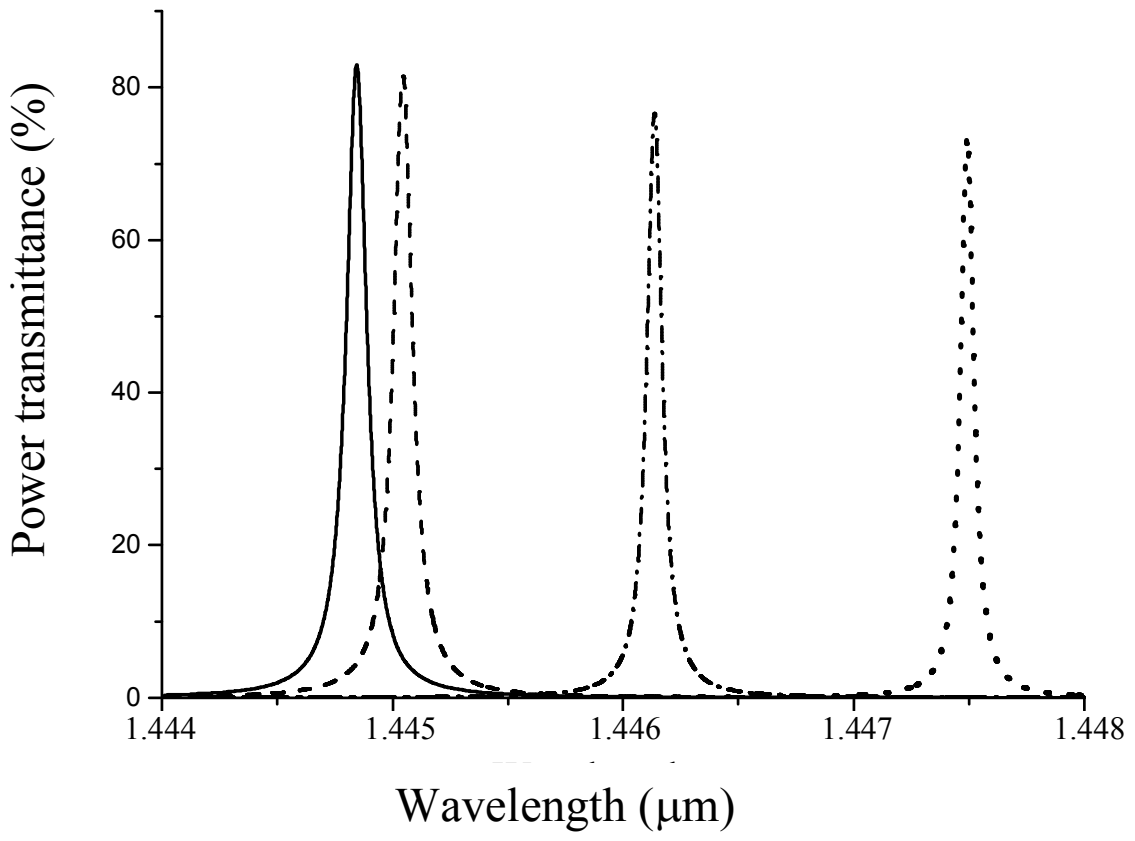

Fig. 7. Transmittance spectra of the p-polarization incident on the MPC of formula $(NM)^{31}(MN)^{18}(NM)^{31}$ at the 50° angle for four different magnetic layers gyrations: $g = 0$ (solid line), $g = 0.01$ (dashed line), $g = 0.03$ (dash-dotted line), and $g = 0.05$ (dotted line).

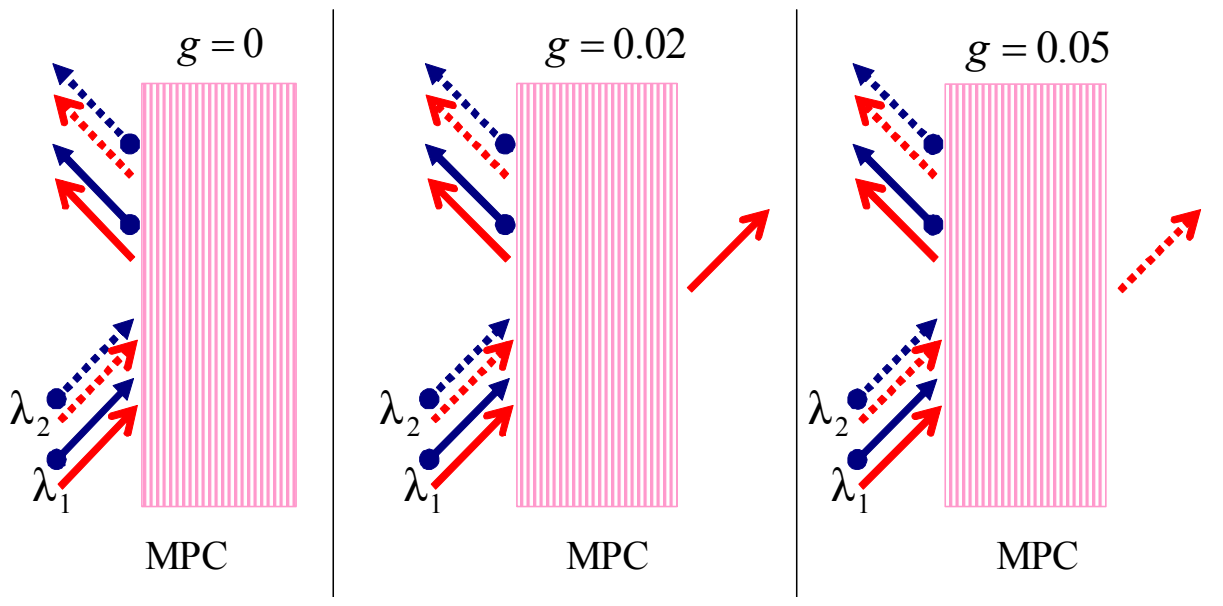

Fig. 8. Schematic diagram of a tunable polarization-dependent optical switch utilising magnetic control of the spectral localization of MPC bandgap defect modes. Depending on the medium gyration, only one polarization type (p-pol. and s-pol. are shown by arrows with circle-ends and simple ends, respectively) of two incident wavelengths (shown by solid and dotted lines) is transmitted for each of the shown gyrations.

The linear shift in the defect mode position has the same nature as the magnetic field induced shift of the band gap edges theoretically predicted and explained in [14]. It is worth noticing that the peaks of different polarization (s-polarization for the considered case) generally occur at different wavelengths. That is why the effect described here is of practical value because it can be utilized for tunable optical devices such as polarizers and switches. The latter, for example, being organized on the basis of the suggested structures will enable to separate s- and p-incident polarizations at the desired wavelength which can be finely tuned by the magnetic field (Fig. 8).

While MPCs have been proposed for optical polarizers, the output light ellipticity can significantly limit the performance of MPC-based polarizers. In the case of the discussed MPC scheme the ellipticity increases with the gyration and it is about 1.5° for the Ce:YIG saturation magnetization corresponding to $g = 0.009$. Consequently, for the best performances, optimization of the MPC structure is vital.

## 4. CONCLUSION

In this paper we considered magnetooptics of 1D MPCs paying much attention to their possible applications for the modern nanoscaled optics devices. For the theoretical description we utilized a transfer matrices formalism adopted to the case of oblique light incidence. Tunability of the band-gap defect modes spectral location by external magnetic fields was found. Several possibilities of one-dimensional magnetic photonic crystals applications for the optical switch and WDM demultiplexers with transmission loss equalization were discussed. Initial steps towards the practical implementation of the proposed devices were reported.

## 5. REFERENCES

[1] E. Yablonovitch, *Phys. Rev. Lett.*, vol. 58, p. 2059, 1987.
[2] V. Berger, *Phys. Rev. Lett.*, vol. 81, p. 4136, 1998.
[3] M. Inoue, T. Fujii, *J. Appl. Phys.*, vol. 81, p. 5659, 1997.
[4] M.J. Steel, M. Levy, R. M. Osgood, *IEEE Photon. Technol. Lett.* vol. 12, p. 1171, 2000.
[5] V.I. Belotelov, A.K. Zvezdin, *JOSA B*, vol. 22, p. 286, 2005.
[6] I.L. Lyubchanskii, N.N. Dadoenkova, M.I. Lyubchanskii, *et al.*, *J. Phys. D: Appl. Phys.* vol. 36, p. R277, 2003.
[7] K. Sakoda. Optical properties of Photonic Crystals. (Springer, 2001).
[8] Amnon Yariv, Pochi Yeh, Optical Waves in Crystals: Propagation and Control of Laser Radiation. (Wiley, 1983) p. 608.
[9] S. Visnovsky *J. Magn. Soc. Jap.* 15, Suppl. S1, p. 67, 1991.
[10] A. Zvezdin, V. Kotov. Modern Magnetooptic and magnetooptical materials. IOP Publishing, Bristol and Philadelphia, p.363 (1997).
[11] M. Vasiliev, V.I. Belotelov, V.A. Kotov, A.K. Zvezdin, K. Alameh, "Magnetic Photonic Crystals: 1-D Optimization and Applications for the Integrated Optics Devices,"*IEEE J. Lightwawe Technol,* to be published, 2005.
[12] Ahderom, S. and Raisi, M. and Alameh, K.E. and Eshraghian, K.,"Dynamic WDM equalizer using Opto-VLSI beam processing," *IEEE Photonics Technol. Lett.,* vol. 15, pp. 1603-1605, 2003.
[13] Rong Zheng, Zhenglin Wang, Kamal E. Alameh, and William A. Crossland. "An Opto-VLSI Reconfigurable Broad-Band Optical Splitter". *IEEE Photonics Tech. Lett.,* vol. 17, p. 339, 2005.
[14] A.K. Zvezdin, V.I. Belotelov, *Euro Phys. J.* B, vol. 37, p. 479, 2004.